\documentclass[twocolumn,prc,aps,floatfix]{revtex4}
\usepackage[dvips]{graphicx}
\begin{document}
\title{Exchange interaction radically changes behavior
 of a quantum particle in a classically forbidden region: simple model}
\author{V.V. Flambaum}
\affiliation{
 School of Physics, The University of New South Wales, Sydney NSW
2052, Australia
}
\date{\today}
\begin{abstract}
Exchange interaction strongly influences the long-range behavior of
localized electron orbitals and quantum tunneling amplitudes.
In the Hartree-Fock approximation the exchange produces a power-law decay
 instead of the
usual exponential decrease at large distances.
To show that this effect is real (i.e. not a result of the approximation)
 we consider a simple model
where different effects may be accurately analyzed. Applications include huge
enhancement of inner electron ionization by a static electric field
or laser field considered in Ref. \cite{Amusia}.
\end{abstract}
\maketitle
%\pacs{PACS: 31.25.-v, 31.25.Eb, 31.25.Jf}
%\pacs{PACS: 06.20.Jr, 33.20.Bx }

\section{Introduction}
  One of the first famous results of Quantum Mechanics was that a particle
 may tunnel through a potential barrier. The tunneling amplitude
is exponentially small in the classical limit.
This result may be incorrect if we take into account
the exchange interaction and correlation effects.
In the Hartree-Fock equations the exchange interaction is described by
 the non-local (integration) operator, and the well-known theorems proven
 for the Schroedinger equation with a local potential $U({\bf r})$ are violated
if we add the exchange term (or any other non-local operator).
According to \cite{Flambaum2009} (see also \cite{DFS,Froese,Handy})
the exchange can produce a power-law decay instead of the
usual exponential decrease at large distances. For inner orbitals inside
 molecules decay is $r^{-2}$, for macroscopic systems
 $\cos{(k_f r)} r^{-\nu}$, where $k_f$ is the Fermi momentum and 
   $\nu=3$ for 1D, $\nu=$3.5 for 2D and $\nu=$4 for 3D crystal.
%Correlation corrections do not change these conclusions.
 Slow decay increases the spin-spin interaction
 between localized
 spins in solids and the under-barrier tunneling amplitudes. 

A very interesting manifestation of this phenomenon has been suggested by Amusia
 in Ref. \cite{Amusia}. He showed that the exchange interaction
may increase probability of ionization of inner atomic electrons by an
 external electric field by many orders of magnitude (in one of the examples
the enhancement factor was $10^{39}$!). Amusia claimed that this enhancement
may explain experimentally observed enhancement of multi-electron ionization
by a strong laser field, see e.g. Ref. \cite{experiment}
 (a different explanation,
an ``atomic antenna'' mechanism, was suggested by Kuchiev \cite{Kuchiev}
and rediscovered by Corkum \cite{Corkum}).

All theoretical results \cite{Flambaum2009,DFS,Froese,Handy,Amusia}
mentioned above have been obtained in the
 Hartree-Fock approximation. Naturally, one may ask a question:
is this enhancement real or is it just an artefact of an  approximate
solution? Indeed, the Hartree-Fock  method ignores correlation
 effects which sometimes play a very important role.
% (for example, the correlations produced by  the strong  electron repulsion
% create the Mott insulator states in crystals).
 An estimate of the correlation effects
have been done in Ref. \cite{Flambaum2009}. The conclusion is that if the
 correlation corrections may be treated using perturbation theory,
 their long-range effect is less significant than that of the exchange.
 However, it is important to consider
 a simple model where different effects can be accurately analyzed and check
if the enhancement of the quantum tunneling by the exchange interaction
really takes place.

Let us consider a  model of resonance tunneling from one potential
well to another potential well. The case of symmetric double-well potential
has been solved e.g. in the textbook \cite{Landau}. There are two levels
corresponding to the symmetric (ground state) and antisymmetric wave
 functions. The tunneling produces the splitting of these levels,
 $E_{\pm}=E_1 \mp t_1$ where
 $t_1 \sim \exp{(-\int |p|dr/\hbar)}$ is the tunneling amplitude,
 $|p|=\sqrt{2m(U(r)-E_1)}$ is the semiclassical under-barrier momentum
 and the integral is taken between the classical turning points.

 If the first  potential well  (``a'') is slightly deeper than the second
potential well (``b''),  the ground state wave
 function may be presented as $\psi_g=\psi_{1a}+B_{t1} \psi_{1b}$ where
$B_{t1} \sim t_1/(E_{1a}-E_{1b})$.
Here we assume that the distance to other levels is large,
 $t_1 \ll (E_{1a}-E_{1b})$ and the probability of the
 particle in the ground state to be in the well $b$ is exponentially small
(proportional to the squared tunneling amplitude,
 $B_{t1}^2 \sim t_1^2/(E_{1a}-E_{1b})^2$).

 Now we add a second particle  (identical fermion or boson) to a higher state 2
 which has energy
close to the top of the barrier.  We can present its wave function as
$\psi_2=A_2\psi_{2a}+B_2\psi_{2b}$ where the coefficient $B_2$ is not
 necessarily small.
In this case the probability
 of the particle in the ground orbital to be in the potential well $b$ is no
 longer proportional  to the exponentially small parameter $t_1^2$. Indeed,
the following two-step process takes place.

 Step 1: the second particle
  tunnels from the potential
well $a$ (orbital $\psi_{2a}$) to the potential well b (orbital $\psi_{2b}$).

Step 2: two-body process $2b,1a \to 1b,2a$ due to  a nondiagonal Coulomb
 exchange interaction  which transfers the first  particle  from orbital
1$a$ to the orbital $2a$ and the second particle  from $2b$ to  $1b$.

 As a result of these two
 steps, we have no change
in the occupation of the state 2 and transfer of  a particle from the ground
 state $1a$ to $1b$.
This gives the amplitude for the ground state particle to be in the well 
``b'':
\begin{equation}\label{BG}
 B_{G1} \sim \frac{G(2,1a;1b,2)}{E_{1a}-E_{1b}},
\end{equation}
 where 
\begin{equation}\label{G}
G(2,1a;1b,2) = \int \psi_{2}({\bf r})^\dagger\psi_{1a}({\bf r})
 \frac{e^2}{|{\bf r-r'}|} \psi_{1b}({\bf r'})^\dagger \psi_2({\bf r'})
 d {\bf r'}d {\bf r}
\end{equation}
is the Coulomb exchange integral. Note that the potential wells here may
have one, two or three dimensions.

This result may also be derived from the Hartree-Fock equation
for the orbital $ \psi_{1}= \psi_{1a}+ \delta \psi_{1}$,
\begin{equation}\label{HF}
-\frac{\hbar ^2}{2m}\frac{d^2}{d{\bf r}^2} \psi_{1}({\bf r}) 
+(U({\bf r})-E_1) \psi_{1}({\bf r})= K({\bf r}),
\end{equation}
by  projecting it to the orbital $ \psi_{1b}$. Here
\begin{equation}\label{HFK}
 K({\bf r})=\psi_2({\bf r}) \int \psi_2({\bf r'})^\dagger
 \frac{e^2}{|{\bf r-r'}|} \psi_{1}({\bf r'}) d {\bf r'}
\end{equation}
is the exchange term. Note that the contribution of the direct
term in the Coulomb interaction between the particles 1 and 2 is included into
 the mean field  potential $U({\bf r})$.

 Equation (\ref{G}) gives us dependence of the amplitude 
$B_{G1}$ on the distance $|a-b|$ between the wells $a$ and $b$.
If the distance $|a-b| \gg r_1$ where $r_1$ is the size of the orbital $1a$,
we can expand  $1/|{\bf r-r'}|$ near $|{\bf r-r'}|=|a-b|$.
Integral with the first term $1/|a-b|$ of this expansion vanishes due to the
 orthogonality of the wave functions $\psi_{1a}({\bf r})$ and
 $\psi_{2}({\bf r})$. Therefore, the expansion starts from $1/|a-b|^2$.

Now we may discuss a contribution of the correlation effects. They correspond
to higher orders in the perturbation theory in the  Coulomb  interaction
 integrals G, so they decay with distance faster than  $1/|a-b|^2$.  

 Similarly, the enhancement of the tunneling  takes place for
the ionization of an inner atomic electron by an external electric field.
 We just need to make the length of the potential well ``b'' infinitely large,
 so the orbitals 1$b$ and 2$b$  will  be in the continuum.

  It is instructive to compare the exchange enhancement mechanism with
the atomic antenna mechanism \cite{Kuchiev,Corkum}. In the antenna mechanism an
external electron is ionized by a strong laser field, oscillates in this field
and accumulates energy. Then this electron collides with the parent ion and
 ionizes it. In the exchange mechanism the external electron plays a passive
role, it does not change its initial state. This may give an additional
 coherent enhancement if the number of  electrons in an external subshell is
 large. Indeed, in many-electron atoms the exchange term in the Hartree-Fock
 equation
(\ref{HF}) contains sum over all electron orbitals,
\begin{equation}\label{HFK1}
 K({\bf r})=\sum_q\psi_q({\bf r}) \int \psi_q({\bf r'})^\dagger
 \frac{e^2}{|{\bf r-r'}|} \psi_{1a}({\bf r'}) d {\bf r'}
\end{equation} 
Therefore, all external electrons
contribute coherently into the effective tunneling amplitude for an 
inner electron. Ref.  \cite{Amusia} claims that this may give an additional
enhancement factor $N_{ext}^2$ in the probability of the ionization
 where $N_{ext}$ is the number
 of external electrons (this dependence  $N_{ext}^2$ is probably  observed
 in the ionization of noble gas clusters 
 in Ref.\cite{Borisov} - see discussion in  Ref.  \cite{Amusia}).  Note,
 however, that different subshells may
 contribute to this  sum in Eq. (\ref{HFK1}) with different signs,
 so the interference is not completely constructive.
Consider, for example, the ionization of $1s$ electron, $\psi_{1a}=\psi_{1s}$.
The sign of the integrals in Eq. (\ref{HFK1}) is determined by the sign
of $\psi_q({\bf r})$ near the origin where $\psi_{1s}$ is located. The large
distance behavior of the corresponding term in $ K({\bf r})$ is determined
by the $\psi_q({\bf r})$ which stays outside the integral. Therefore,
the sign of the contribution of a sub-shell depends on the number of radial
oscillations of the wave function $\psi_q({\bf r})$ which determines the sign
of the product $\psi_q$(near zero)$\psi_q$(outside the atom).
One should take into account this fact when estimating the coherence
 enhancement factor $N^2$.

    In the discussion above we assumed that the residual Coulomb interaction
(beyond the mean field) between the electrons is sufficiently small to be
 treated perturbatively. We should check if the exchange enhancement survives
 in the case of a stronger Coulomb interaction. Consider the two-well problem
 with a very large Coulomb repulsion between two particles. A minimum of the
 Coulomb energy
is achieved when the particles are in different wells, in  state
$\psi_{1a}\psi_{2b}$ or  $\psi_{1b}\psi_{2a}$. Mixing between these states
may be produced by the non-diagonal exchange interaction $G(2b,1a;1b,2a)$,
i.e. it does not require any tunneling at all. Two other states
$\psi_{1a}\psi_{2a}$ and  $\psi_{1b}\psi_{2b}$ are separated from the lower
states by the large Coulomb energy  $Q=Q_{aa}-Q_{ab}$ where $Q_{aa}$ and
 $Q_{ab}$ are the Coulomb energies for the particles in the same well
 and different wells correspondingly.  Mixing between the states
$\psi_{1a}\psi_{2a}$ and  $\psi_{1b}\psi_{2b}$ may be achieved in 3 steps.

Step 1: tunneling of particle from $2a$ to $2b$ with creation of an
 intermediate state $\psi_{1a}\psi_{2b}$ separated by the energy interval $Q$.

Step 2: the non-diagonal exchange interaction $G(2b,1a;1b,2a)$ transfers     
$\psi_{1a}\psi_{2b}$ to $\psi_{1b}\psi_{2a}$.
 At this step we have mixing of the single-particle states 1$a$ and 1$b$,\\
$B_{G1} \sim t_2 G(2b,1a;1b,2a)/Q^2$.

 Step 3: the tunneling from $2a$ to $2b$.

 We see again that we do not need the exponentially small
tunneling amplitude $t_1$, i.e. the exchange enhancement works.
The only suppression we have here is due to the large Coulomb
matrix element $Q$ in the denominator of the mixing amplitude\\
$t_{eff} \sim t_2^2 G(2b,1a;1b,2a)/Q^2$\\
(Note that a similar suppression due to
a large value of $Q$ transforms a half-filled conducting band in solids into
 the Mott insulator. This transition influences the exchange power tail
 for a localised electron in solids - see discussion in \cite{Flambaum2009}).

 Similar results may be obtained for an attraction between
 the particles. This may be a model for a tunneling of an inner
electron through a Josephson junction.
 
Thus,  the exponential enhancement of the tunneling due to the exchange
interaction really exists.

This work is  supported by the Australian Research
Council. I am grateful to J. Berengut for useful comments.

\end{document}